\documentclass[prb,twocolumn,superscriptaddress,showpacs,amsmath,amssymb]{revtex4}

\usepackage{graphicx}
\usepackage{latexsym}
\usepackage{amsmath}
\usepackage{amssymb}
\usepackage{amsfonts}
\usepackage{color}
\usepackage{bm}
\usepackage{verbatim}
\usepackage{color}

\begin{document}
\title{Majorana fermions on the Abrikosov flux lattice in a $p_x+ip_y$
superconductor and thermal conductivity in superclean regime}

\author{M.~A.~Silaev}
\affiliation{Institute for Physics of Microstructures, Russian
Academy of Sciences, 603950 Nizhny Novgorod, GSP-105, Russia}

\date{\today}

%%% abstract
\begin{abstract}
We show that periodic lattice of Abrikosov vortices in chiral $p$
wave superconductor in general supports fermionic states with zero
energy provided the intervortex distance is smaller than the
critical one.  The zero modes appear at the intersection with the
Fermi level of electronic magnetic Bloch bands formed by the
overlapping vortex core states. The Bloch bands are robust against
lattice disorder induced by fluctuations of vortex positions and
can transmit the energy flow across the lattice. The hallmark of
zero modes on Bloch bands in electronic heat conductivity is
discussed.
\end{abstract}

\maketitle

Topological 2+1 dimensional Fermi systems are one of the most
intriguing topics in the field of condensed matter physics. An
intense investigation of such systems has started from the theory
of integer quantum Hall effect when it was shown that the
transverse conductivity is proportional to the discrete valued
topological invariant of the ground state \cite{IQHE} related to
the first Chern number of the Berry phase gauge field in the
Brillpuin zone. An analogous topologically nontrivial ground state
was found in superfluid A phase of $^3$He films\cite{VolovikHe3}.
In this system the non-trivial chiral $p_x+ip_y$ structure of
superfuid order parameter corresponding to the Cooper pairing with
angular momentum $L_z=\pm 1$ allows for the existence of quantum
Hall effect in the absence of magnetic field. The same state is
suggested to realize in layered triplet $p$ wave superconductor
Sr$_2$RuO$_4$ \cite{SrRuO}.

Many of the exotic properties of chiral $p_x+ip_y$ superconductors
and Fermi superfluids are determined by the interplay of the
ground state topology and the properties of fermionic bound states
modified in the vicinity of topological defects in order parameter
distribution.  In particular the fermionic sectors of $^3$He A and
Sr$_2$RuO$_4$ contain zero energy states localized near domain
walls and solitons\cite{SilaevFermiArcs},
boundaries\cite{Tsutsumi} and quantized
vortices\cite{VolovikPwave}. The zero energy fermionic modes can
be described in terms of the self-conjugated Majorana fermions
which were theoretically predicted to appear in several other
two-dimensional systems such as the fractional quantum Hall liquid
at filling $5/2$ \cite{FQHE}, heterostructures of topological
insulators and superconductors \cite{Heterostructures}, and
possibly certain Iridates which effectively realize the Kitaev
honeycomb model \cite{Irridates}.
  %%%%%%%%%%%%%%%%%%% Fig 1 %%%%
  \begin{figure}[t]
  \centerline{\includegraphics[width=1.0\linewidth]{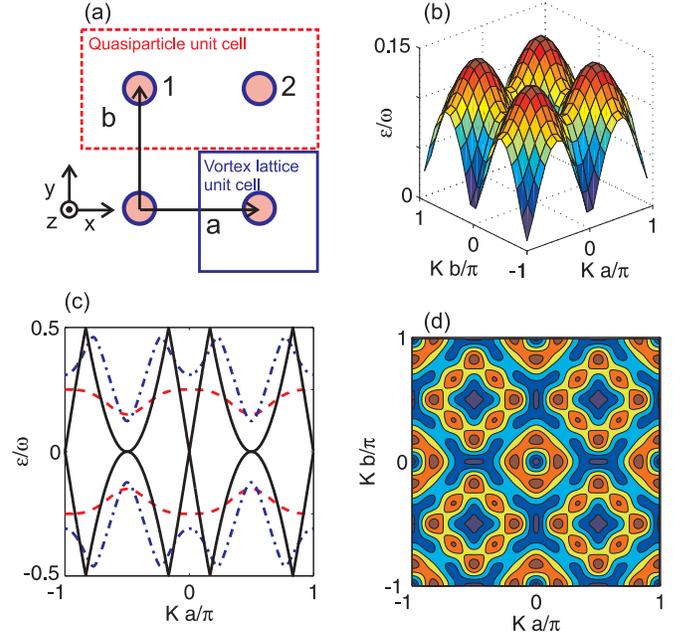}}
  \caption{\label{Fig1} (a) The vortex lattice unit cell and corresponding quasiparticle
  unite cell containing the nodes marked by 1 and 2 in square vortex lattice.
  The vortex positions are marked with red filled circles which form the
  2D Brave lattice with the basis $({\bf a , b})$.
  (b) The 3D plot of magnetic Bloch band $\varepsilon_{0+} ({\bf
  K})$ given by (\ref{Eq:spectrumLattice}) $\chi=\pi/4$ and
  $I_{\bf a}=I_{\bf a}= 0.14$. (c) The Bloch bands $\varepsilon_{0\pm} (K_x,K_y=0)$
   for $I_{\bf a}=I_{\bf b}= 1.4 ;\; 2.8;\; \sqrt{2}\pi $
   shown by red dashed, blue dash-dotted and black solid lines correspondingly.
   (d) The contour plot of $\varepsilon_{0+}({\bf K})$
   for  $\chi=\pi/4$ and $I_{\bf a}=I_{\bf b}= 4.3$.
   The plots (b,c,d) correspond for the square vortex lattice. }
  \end{figure}

An appealing possibility offered by the nontrivial structure of
fermionic spectrum in the vortex phase of chiral $p_x+ip_y$
superconductors is the realization of quantum matter with exotic
non-Abelian quasiparticle statistics \cite{SternNature,
IvanovPRL}.
 In this case the non-Abelian anyons are
 presented by vortex excitations supporting zero-energy Majorana
 fermions residing inside their cores. That is the spectrum of
vortex core fermions is given by
\begin{equation}\label{Eq:CdGM}
\varepsilon= \omega (n+\gamma)
\end{equation}
where $n$ is integer number, $\gamma=1/2$ for $s$ wave \cite{CdGM}
and $\gamma=0$ for $p_x+ip_y$ wave \cite{VolovikPwave}
superconductors. Thus in topologically non-trivial superconductors
the spectrum of vortex core fermions (\ref{Eq:CdGM}) contains
zero-energy modes with $n=0$ which can be conveniently described
in terms of Majorana self-conjugated fermionic
field\cite{IvanovPRL}.  The ground state of the system with
multiple spatially well separated vortices with zero bound
fermionic modes is topologically degenerate.  The non-Abelyan
statistics of vortex anyons allows for the unitary transformations
of the ground state realized through the adiabatic permutation of
vortices. Such possibility provides an extra motivation for the
study of vortices in $p_x+ip_y$ superconductors due to their
potential application in topological quantum computing
\cite{TQCrmp}.

Vortex core Majorana fermions have an important property of being
stable with respect to the impurity scattering\cite{VolovikPwave}
and order parameter perturbations \cite{IvanovPRL}. However the
spectrum vortex core states is extremely sensitive to the
intervortex quasiparticle tunnelling. The corresponding spectrum
modification in finite clusters of vortices was investigated first
by Mel'nikov and Silaev \cite{MS1,MS2}. In particular for the
generic problem of two vortices placed at the distance $d$ the low
energy fermionic spectrum has the form
 \begin{equation}
 \varepsilon=\frac{\omega}{\pi}\left[ \pm \arccos(
  \sqrt{1-e^{-\alpha}}\sin\beta )+
 \pi(n+1/2+\gamma ) \right]\, \label{Eq:spectrum2v}
 \end{equation}
 where $\gamma=1/2$ for $s$ wave and $\gamma=0$ for $p_x+ip_y$ wave superconductivity,
  $\alpha = 2\pi e^{-2d/\xi} k_F\xi^2/d$, $\beta =  k_F d +\alpha
 \ln(k_F\xi^2/d)+\arg(\Gamma(1-i\alpha))+\pi/4$ where $\Gamma (x)$
 is Gamma function.
  The spectrum (\ref{Eq:spectrum2v}) contains two series of levels
  with the interlevel distance $\omega\simeq \Delta_0/(k_F\xi)$, where $\Delta_0$ is the gap
value far from the vortex core, $\xi$ is the superconducting
coherence length.
 The Eq.(\ref{Eq:spectrum2v}) demonstrates that
 the perturbation of energy levels due to the intervortex
 quasiparticle tunnelling in general is not described by the plain
 tight binding theory.
 Indeed the shift of energy levels with respect to the isolated
 vortex spectrum becomes larger than the energy level spacing
 $\omega$ when the intervortex distance is smaller than the critical one
  $d<d_c$, where $d_c \simeq (\xi/2) \ln(k_F\xi)$ is much larger
 then $\xi$ since $k_F\xi\gg 1$. For the typical parameter $k_F\xi =100$ corresponds to
  magnetic fields being larger than that of the order $0.1
  H_{c2}$.

Thus in a pair of vortices the intervortex quasiparticle
tunnelling removes the twofold degeneracy of vortex core states.
In particular in $p_x+ip_y$ superconductor it splits the Majorana
zero energy states provided the phase $\beta\neq \pi n$ in the
Eq.(\ref{Eq:spectrum2v}). Such splitting of zero energy levels
opens the gap in the fermionic spectrum and can break the quantum
coherence during the vortex permutation which is important for the
fault tolerance of topological quantum computations
\cite{Galitskii}.
 The generalization of Eq.(\ref{Eq:spectrum2v}) for $M-$ vortex
clusters is straightforward and was discussed in
detail\cite{MS1,MS2}. In particular for an odd number of vortices
$M$ there is always at least one zero energy state irrespective of
the vortex position in the cluster. Thus the splitting of Majorana
fermions in vortex cluster is not a generic effect and depends on
the parity of the total number of vortices $M$. On the other hand
in clean type-II superconductors without disorder and pinning
centers vortices in finite magnetic field form periodic Abrikosov
flux lattice. Therefore the natural question considered in the
present paper is whether the spectrum of fermions on the vortex
lattice in chiral $p_x+ip_y$ superconductor is gapped or contains
zero energy Majorana states.

Previously the various types of two dimensional lattice spectrum
problems of Majorana fermions were considered
\cite{Kitaev,Laumann,SternSquare,SternNJP,Lahtinen}.
 These models take into account only the tunnelling between lowest energy states in the vortex cores.
As we have discussed above in the generic problem of two vortices
the shift of vortex core energy levels becomes larger than the
interlevel energy already at small magnetic fields $H\geq 0.1
H_{c2}$. Therefore the one level approximation of lattice models
is applicable for sparse vortex lattices. Instead in the present
paper we consider the eigenvalue problem of genuine Bogolubov - de
Gennes equation in chiral $p_x+ip_y$ superconductor with gap
potential corresponding to the periodic Abrikosov flux lattice. To
treat this problem we generalize the original approach developed
earlier \cite{MS1,MS2} to calculate the fermionic spectra of
finite vortex clusters. This approach allows to calculate the
spectrum when the intervortex distance is $d\geq \xi$.

In general the problem of identifying the quasiparticle energies
in superconductors is to solve the Bogolubov - de Gennes (BdG)
equations having the form:
  \begin{equation}\label{Eq:BdG}
\hat H_0\Psi +\left(0\quad \;\,\hat \Delta\atop \hat \Delta^+\quad
0\right)\Psi =\varepsilon\Psi \ ,
\end{equation}
 where $ \hat H_0=\hat\tau_3\left[(\hat {\bf p}- \hat\tau_3{\bf
 A})^2-k_{F}^2 \right]/2m $,
 $\Psi=(U,V)$, $U$ and $V$ are the particle- and hole - like parts of the fermionic quasiparticle wave
 function,
 $\hat\sigma_i$ are Pauli matrices,
 $\hat {\bf p}=-i\nabla$,
 $\hat\tau_i$ are the Pauli matrices in a particle--hole space,
 the gap operator is
 $ \hat\Delta=\left\{\Delta
 (\hat{\bf{r}}),e^{i\chi\theta_p}\right\}$
  where $\chi=\pm 1$ is chirality, $\hat{\bf{r}}$ is a coordinate
operator,
 $\Delta ({\bf{r}})$ describes the spatial dependence of the
 gap function and $\{A,B\}=(AB+BA)/2$ is an anticommutator which
 provides the gauge invariance of $\hat\Delta$. The  phase of the order parameter depends on the direction
 of the electron momentum in $xy$ plane: ${\bf p}=p
(\cos\theta_p,\sin\theta_p)$.
 %Here we omit the spin--dependent part of the gap operator $\hat
%\Delta$, neglecting the spin--orbit interaction. Also we neglect
%the dispersion of QP energy in the direction perpendicular to the
%anisotropy plane $xy$, assuming a cylindrical Fermi surface.
  The
magnetic field is directed along the $z$ axis ${\bf B}=B{\bf z}$
and for extreme type-II superconductors we can consider the
magnetic field to be homogeneous on the spatial scale of
intervortex distance and take the gauge
 ${\bf A}=[{\bf B}\times {\bf r}]/2$ where ${\bf B}$ is an average magnetic
 field.

%%%%%%%%%%%%%%%%%%%%%%%%%%%%%%%%%%%%%%%%%%%%%%%%%
 Then the periodicity of vortex lattice is determined
\begin{eqnarray}\label{Eq:VLperiodicity}
\Delta ({\bf r+d})=e^{i{[\bf B\times d]  r}+i\varphi_{\bf d}}
\Delta ({\bf r}) \\ {\bf A} ({\bf r+d})={\bf A} ({\bf r})+[{\bf
B}\times{\bf d}]/2
\end{eqnarray}
where ${\bf d}=n_a {\bf a}+n_b {\bf b}$ is the translation of the
vortex lattice, $n_a,n_b$ are integer numbers and $\varphi_{\bf
d}$ is an arbitrary constant phase shift. By choosing the
Wigner-Seitz elementary cell of vortex lattice and placing the
origin ${\bf r}=0$ at the vortex center in this cell we
immediately obtain that $\varphi_{\bf a}=\varphi_{\bf b}=\pi$.

The translational properties of $\Delta ({\bf r})$ and ${\bf
A}({\bf r})$ make the Eq.(\ref{Eq:BdG}) to commute with the
magnetic translation operator
\begin{equation}\label{Eq:MagneticTranslation}
T^h_{{\bf d}}= \hat\tau_3 e^{i\hat\tau_3{[\bf B\times d] r}/2}
T_{{\bf d}}
\end{equation}
so that $T^h_{{\bf d}} \hat H T^{h+}_{{\bf d}}= \hat H$
 where $T_{{\bf d}}$ is the usual translation by the lattice
 vector ${\bf d}$. Consequently the solutions of the BdG Eq.(\ref{Eq:BdG})
 can be classified according to the eigenstates of magnetic
 translation operator. An important point is that the magnetic flux through the
 vortex lattice unit cell is one half of the flux quantum
 $B {\bf z}\cdot  [{\bf a \times \bf b}] = \pi$ so that the magnetic translations by lattice vectors
 anticommute $T^h_{{\bf a}}T^h_{{\bf b}}=-T^h_{{\bf b}}T^h_{{\bf a}}$. Therefore we
 should introduce the unit cell for the quasiparticle functions
 consisting of two vortex lattice unit cells, for example shifted by the vector ${\bf a}$.
 For the case of square vortex lattice this choice is illustrated
 in the Fig.\ref{Fig1}(a). Then the magnetic translation subgroup is formed by vectors
 ${\bf d_m}=2n_a {\bf a}+n_b {\bf b}$ and the solution of
 Eq.(\ref{Eq:BdG}) in general has the form
\begin{equation}\label{Eq:BlochWave}
\Psi=\sum_{\bf d_m} e^{i{\bf K  d_m}} T^h_{{\bf d_m}} \left[\Psi_1
({\bf r}) +  e^{i{\bf K  a}} \Psi_2 ({\bf r})\right]
\end{equation}
where the functions $\Psi_{1,2} ({\bf r})$ are localized in the
centers of vortices forming the unit lattice cell for the
quasiparticles (see Fig.\ref{Fig1}). We substitute the ansatz
 (\ref{Eq:BlochWave}) to the BdG Eq.(\ref{Eq:BdG}) and calculate the
 inner product with $\Psi_{1,2}({\bf r})$ taking into account only
 overlap with neighbor cites to obtain the system of tight binding
 equations.

 The further calculation requires
 expansion of the node wave functions $\Psi_{1,2} ({\bf r})$ by
 the basis of localized fermionic states of an isolated vortex. It
  can be implemented using the quasiclassical approximation and the convenient
  formalism\cite{MS1,MS2} of the so called
 $s-\theta_p$ representation which allows to
 express the quasiparticle wave function in momentum representation in the form:
 \begin{equation}
 \label{st}
 \Psi({\bf p})=
 \frac{1}{k_F}\int\limits_{-\infty}^{+\infty} ds
 e^{-i(|{\bf p}|- k_F)s/\hbar}
 \psi(s,\theta_p) \ .
 \end{equation}
 The equation for $\psi(s,\theta_p)$ reads: $\hat H\psi =E
 \psi$, where
 \begin{equation}
 \label{bdg} \hat H = -i V_F
 \hat\tau_3\partial_s+ \left(0\quad \;\,\hat \Delta\atop \hat \Delta^+\quad
0\right)  \ ,
 \end{equation}
 $V_F= k_F/m$ is Fermi velocity. Here we take into
 account the quantization of angular motion variable by treating
 the angular momentum as differential operator $\hat\mu =-i\partial/\partial \theta_p$.
 Hence the spatial coordinate in the gap operator in Eq.(\ref{bdg})
 is quantum variable in $s-\theta_p$ representation
 $ \hat {\bf r} =s {\bf k}_F/k_F +
 \left\{ [{\bf k}_F, {\bf z}], \hat\mu\right\}/k_F^2$.
Let us emphasize that the Hamiltonian (\ref{bdg}) takes account of
noncommutability of $\hat\mu$ and $\theta_p$ and, thus, the above
description involves the angular momentum quantization. Replacing
$\hat\mu$ by a classical variable we get Andreev equations along
straight trajectories. The inner product can be expressed through
the envelope functions
\begin{equation}\label{Eq:InnerProduct1}
\langle \Psi_1|\Psi_2\rangle =
\frac{\pi}{k_F}\int_{-\infty}^{\infty} ds\int_0^{2\pi}d\theta_p
 \psi^+_{1}\psi_{2} (s,\theta_p)
\end{equation}
 and the magnetic translation operator (\ref{Eq:MagneticTranslation}) in $s-\theta_p$
 representation has the form
 %\begin{equation}\label{Eq:MagneticTranslationST}
% T^h_{\bf d}=\tau_3
% e^{-i\varphi_h({\bf d})\tau_3/2} T_{\bf d}
% \end{equation}
$ T^h_{\bf d}=\hat\tau_3 e^{-i\varphi_h({\bf d})\hat\tau_3/2}
T_{\bf d}$ where $\varphi_h({\bf d}) = d (s+{\bf nd})
\sin(\theta_p-\theta_d)$, the angle $\theta_d$ defines the
direction of ${\bf d} = d(\cos\theta_d, \sin\theta_d)$
 and $T_{\bf d}=\exp\left[-i{\bf d  k_F}
 (1-ik_F^{-1}\partial_s)\right]$ is the translation
 operator\cite{MS1}.

The form of the node functions $\Psi_{1,2}(\bf r)$ in
Eq.(\ref{Eq:BlochWave}) is determined by the states localized in
isolated vortex. We consider the vortex lattice cite $1$ centered
at the origin ${\bf r}=0$ and
 define the spatial dependence of gap function inside the unit cell
  as $\Delta ({\bf r})= \Delta_v (r) e^{i\theta}$
where $\Delta_v (r=0)=0$. Then eigenfunctions of the
 Hamiltonian (\ref{bdg}) centered at lattice cites
 $1$ and $2$ have the form
   \begin{eqnarray}\label{Eq:NodeFunctions}
 \psi_{1}(s,\theta_p)=C_1(\theta_p) \psi_v(s,\theta_p)\\
 \psi_{2}(s,\theta_p)=C_2(\theta_p)  T^h_{\bf a} \psi_v(s,\theta_p)
 \end{eqnarray}
 where ${\bf n} = {\bf k_F}/k_F$ and
 \begin{equation}
 \psi_v (s,\theta_p)= e^{i\hat\tau_3 (1+\chi)\theta_p/2}
 \left(1\atop -i\right)
 \frac{e^{-K(s)}}{\sqrt{\Lambda}} ,
 \end{equation}
 $K(s)=V_F^{-1}\left|\int\limits_0^{s} \Delta_v (t) dt\right|$ and $\Lambda$
 is normalizing factor so that $\langle \Psi|\Psi\rangle =1$ .
  The node functions are translated to other sides according to the
Eq.(\ref{Eq:MagneticTranslation}) so that with the help of
Eq.(\ref{Eq:InnerProduct1}) we obtain the inner products, e.g.
 \begin{eqnarray}\label{Eq:Inner1}
 \langle\Psi_j|\hat H|\Psi_j\rangle= \omega\int_0^{2\pi} C^*_j \hat\mu
 C_j\frac{d\theta_p}{2\pi}\\ \label{Eq:InnerOverlap1}
 \langle\Psi_j|\hat H|T^h_{\bf b}\Psi_j\rangle
 =  i (-1)^{j+1} J_{\bf b}  \int_0^{2\pi} e^{-i{\bf b  k_F}} C^*_j C_j,
\end{eqnarray}
% \begin{eqnarray}\label{Eq:Inner1}
% \langle\Psi_j|\hat H|\Psi_j\rangle= \omega\int_0^{2\pi} C^*_j \hat\mu
% C_j\frac{d\theta_p}{2\pi}\\ \label{Eq:InnerOverlap1}
% \langle\Psi_j|\hat H|T^h_{\bf b}\Psi_j\rangle
% =  i (-1)^{j+1} J_{\bf b}  \int_0^{2\pi} e^{-i{\bf b  k_F}} C^*_j C_j
% \frac{d\theta_p}{2\pi} \\ \label{Eq:InnerOverlap2}
% \langle\Psi_1|\hat H|\Psi_2\rangle= i J_{\bf a} \int_0^{2\pi} e^{-i{\bf a  k_F}} C^*_{1} C_{2}
% \frac{d\theta_p}{2\pi}\\ \label{Eq:InnerOverlap3}
% \langle\Psi_1|\hat H|T^h_{-2\bf a} \Psi_2\rangle =iJ_{\bf a}
%  \int_0^{2\pi} e^{i{\bf a  k_F}} C^*_{1} C_{2}
% \frac{d\theta_p}{2\pi},
%\end{eqnarray}
 $j=1,2$ where ${\bf d= a,b}$ and the Hamiltonian $\hat H$ is given by (\ref{bdg}).
 The sign in Eq.(\ref{Eq:InnerOverlap1}) determined by the
 magnetic flux through the unit cell. Here for simplicity we
 take into account the overlap with four neighboring vortices. The
 cases of more neighbors can be considered analogously.
 % is compensated by the anticommutation of tau_3 from translation and tau_1 from \Delta.
   The main contribution to the inner products of the form
 %(\ref{Eq:InnerOverlap1}, \ref{Eq:InnerOverlap2}, \ref{Eq:InnerOverlap3})
 (\ref{Eq:InnerOverlap1})
 comes from the stationary points of the phases $({\bf a  k_F})$ and $({\bf b  k_F})$ that is
 $\theta_p^*=\theta_{a,b}+\pi n$.
The stationary points $\theta_p^*$ correspond to the trajectories
passing through both of the neighbor vortex cores which means that
we can calculate the overlap factors as follows
 \begin{equation}\label{Eq:overlap}
 J_{\bf d}= \frac{\pi}{k_F} \int_{-\infty}^{\infty}
 \left[\Delta(s)-\tilde{\Delta}_v(s_1) \right]\tilde{\psi}_v^+ (s)\hat\tau_2 \tilde{\psi}_v (s_1) ds.
 \end{equation}
 where $s_1=s-{\bf n d}$,  $\tilde{\Delta}_v(s)=\Delta_v(s) sign (s)$,  and $\tilde{\psi}_v (s)=\psi_v
 (s,\theta_p=0)$.  Then with good accuracy Eq.(\ref{Eq:overlap})
 yields an estimation
 $\left|J_{\bf d}\right|\approx \Delta_0 \exp\left(-d/\xi\right)$.

 With the help of the inner products
 %(\ref{Eq:Inner1},\ref{Eq:InnerOverlap1},\ref{Eq:InnerOverlap2},\ref{Eq:InnerOverlap3})
 (\ref{Eq:Inner1},\ref{Eq:InnerOverlap1})
  %to the Eqs.(\ref{Eq:1},\ref{Eq:2}) and taking
 and taking into account that $J_{\bf d}=-J_{-\bf d}$
  we obtain the equations
 \begin{eqnarray}\label{Eq:C}
 (\varepsilon-\omega\hat\mu) C_1 = F_{\bf b} (\theta_p) C_1 + F_{\bf a} (\theta_p)
 C_2\\ \nonumber
 (\varepsilon-\omega\hat\mu) C_2 = -F_{\bf b} (\theta_p) C_2 + F_{\bf a} (\theta_p) C_1
 \end{eqnarray}
 where $F_{\bf d} (\theta_p) =J_{\bf d} \sin[{\bf d} ({\bf k_F+ K} )]$
 and ${\bf d} = {\bf a,b}$.
The system (\ref{Eq:C}) should be solved together with periodic
boundary conditions
 $C_{1,2}(\theta_p)=C_{1,2}(\theta_p+2\pi)$ for $p_x+ip_y$ wave and $C_{1,2}(\theta_p)=C_{1,2}(\theta_p+2\pi)$ for $s$ wave.
  Note that in Eqs.(\ref{Eq:C})
 we can take into account overlapping with next-to neighbor vortices
  which will introduce the corrections of the relative order
  $e^{-d/\xi}$ to the coefficients $F_{\bf a, b}$. Here we neglect
  such corrections.

%The spectrum of Eq.(\ref{Eq:C}) $\varepsilon=\varepsilon({\bf K})$
%is defined in the magnetic Brilluoin zone which can be chosen as
%$-\pi <{\bf Kb}\leq \pi$ and $-\pi/2<{\bf Ka}\leq \pi/2$. Besides
%it has an additional symmetry $\varepsilon ({\bf K})= \varepsilon
%({\bf K}+{\bf b_r}/2)$ where ${\bf b_r}$ is the reciprocal lattice
%vector  ${\bf b_r}b=2\pi$. This symmetry is the same as for the
%usual tight binding 2D electron model with half quantum flux
%through the unit cell. Moreover taking the complex conjugation and
%changing $\varepsilon\rightarrow -\varepsilon$ we obtain the
%symmetry $\varepsilon ({\bf K})=-\varepsilon ({\bf K})$.

 To solve the Eq.(\ref{Eq:C}) we use the approximate method
 employed earlier for the system of two vortices\cite{MS1}. That
 is besides the vicinity of the angles  $\theta_p^*=\theta_{a,b}+\pi n$
 the solution with good accuracy is $C_{1,2} \sim e^{i \varepsilon
 \theta_p/\omega}$. In the $\delta$ vicinity of the angles $\theta_p^*$ the
 system (\ref{Eq:C}) diagonalizes yielding the matching conditions
 ${\bf C} (\theta_p^*+\delta) = \hat M {\bf C} (\theta_p^*-\delta)$
 for the vector ${\bf C}=(C_1,C_2)^T$. The matching matrices are
 \begin{eqnarray}\label{Eq:MatchingMatrix}
 &\hat M& (\theta_b)=\exp (-i\hat\tau_3\chi_{b+})  \\
 &\hat M& (\theta_b+\pi)=\exp (-i\hat\tau_3\chi_{b-}) \\
 &\hat M& (\theta_a)=\cos\chi_{a+} - i\hat\tau_1 \sin\chi_{a+} \\
 &\hat M& (\theta_a+\pi)=\cos\chi_{a-} - i\hat\tau_1 \sin\chi_{a-}
 \end{eqnarray}
 where
 %$\chi_{d\pm}=\pm (J_{\bf d}/\omega)\sqrt{\pi/k_Fd} \sin (k_Fd-\pi/4\pm {\bf Kd})$.
% $\chi_{d\pm}=I_{{\bf d}c} \sin ({\bf Kd}) + I_{{\bf d}s} \cos ({\bf
% Kd})$ where $I_{{\bf d}c} = (J_{\bf d}/\omega)\sqrt{\pi/k_Fd} $
% and $I_{{\bf d}s} = (J_{\bf d}/\omega)\sqrt{\pi/k_Fd}$
 $\chi_{d\pm}=I_{\bf d}\sin ({\bf Kd} \pm \chi)$ where $\chi=k_Fd-\pi/4$
 and $I_{{\bf d}} = (J_{\bf d}/\omega)\sqrt{\pi/k_Fd}$.

  The periodic boundary condition and Eqs.(\ref{Eq:MatchingMatrix}) yield
   the Bloch waves $\varepsilon=\varepsilon_n({\bf K})$ in a periodic Abrikosov flux lattice
  \begin{equation}
 \varepsilon_{n\pm}({\bf K})=
 \omega \left(\pm\frac{\arccos X}{2\pi} +
  n +\gamma\right) \label{Eq:spectrumLattice}
 \end{equation}
 where   $\gamma=1/2$ for $s$ wave and $\gamma=0$ for $p_x+ip_y$
 wave,
 $X = \cos\chi_{a-}\cos\chi_{a+}\cos(\chi_{b+}+\chi_{b-})-
 \sin\chi_{a-}\sin\chi_{a+}\cos(\chi_{b+}-\chi_{b-})$ and $n$ is
 integer. The width of Bloch bands (\ref{Eq:spectrumLattice}) is
 determined by the overall amplitude $max (\omega, \Delta_0
e^{-d/\xi}/\sqrt{k_Fd})$ and rapid oscillations with the
 period $k^{-1}_F$ by the intervortex distances $a,b$. The phase of oscillations is
 determined by the average magnetic field, e.g.
 $a=b=\sqrt{\Phi_0/B}$ for the square lattice where
 $\Phi_0$ is magnetic flux quantum for Cooper pairs.
 Notwithstanding the rapid oscillations of the bandwidth
 the spectrum (\ref{Eq:spectrumLattice}) survives
  fluctuations of the vortex positions ${\bf \delta r}$
  provided their amplitude is relatively small. Indeed in case of the
  disordered vortex lattices let us search the quasiparticle
  waves in the form $\Psi ({\bf r}) = \Psi_{0} ({\bf r})+ \tilde{\Psi} ({\bf r})$ where the first
  term is periodical and given by Eq.(\ref{Eq:BlochWave}) and the second term
  is the distortion due to the fluctuation of vortex positions. We
  use the expansion (\ref{st},\ref{Eq:NodeFunctions}) with the coefficients $\tilde{C}_{1,2},
  (\theta_p)$ to represent the distortion $\tilde{\Psi} ({\bf r})$ at the particular lattice site.
  Then we get the equations for $\tilde{C}_{1,2}(\theta_p)$
  $(\varepsilon-\omega\hat\mu) \tilde{C}_j = \omega \tilde{\mu}  C_{j0} (\theta_p) $
 where $\tilde{\mu}= {\bf z [\delta r\times k_F]}$ and $C_{j0} (\theta_p)$ corresponds to the
 periodical part of function $\Psi_{0} ({\bf r})$ determined by the solution of Eq.
 (\ref{Eq:C}). The functions $C_{j0} (\theta_p)$ are the rapidly oscillating
 ones with the characteristic period $\Delta \theta _p\approx
  e^{d/\xi}/(k_F \xi)$ so that the amplitude of the distortion
 is small
  $\tilde{C}_j \sim e^{d/\xi}|\delta {\bf r}|/\xi  \ll 1$ provided vortex
  position fluctuations are small enough $|\delta {\bf r}| \ll\xi
  e^{-d/\xi}$.

The plot of the Bloch band $\varepsilon_{0+}({\bf K})$  is
 shown in the Fig.\ref{Fig1}(b) for the parameters $\chi=\pi/4$ and
 $I_{\bf a}=I_{\bf b}=0.14$. One can see that this band contains small energy
 gap. Indeed for $|I_{\bf d}| \ll 1$ the Eq.(\ref{Eq:spectrumLattice})
 can be simplified. Taking into account the quadratic terms of the order $I_{\bf d}^2$
 we obtain the gapless spectrum identical to the one level lattice
 model \cite{SternSquare} with the hopping amplitude determined by $\omega|I_{\bf d}|$.
 On the other hand the terms of the order
 $I_{\bf d}^4$ open the gap in the spectrum which is beyond the
 accuracy of one level approximation and appear due to the
 mixing with higher levels. Note that overlap with next-to
 neighbor vortices introduce correction smaller by the factor
 $1/\sqrt{k_F\xi}$ than the interaction with higher levels.
 Decreasing the intervortex distance one finally obtains gapless
 spectrum when $|I_{\bf d}| \geq \pi$. The example of such crossover to the gapless regime is
 demonstrated in the dependence $\varepsilon (K_x,K_y)$ for the square lattice in Fig.\ref{Fig1}(c).
 We set $\chi=\pi/4$ and plot by red dash-dotted, blue dashed and black solid lines
 the Bloch bands $\varepsilon_{0\pm} ({\bf K})$ for $I_{\bf a}=I_{\bf a}= 1.4;\; 2.8;\; \sqrt{2}\pi$
 correspondingly. For higher values of the overlap $|I_{\bf d}|$ the structure of Bloch
 bands becomes more complicated with rapid oscillations as
 function of quasimomentum with the characteristic period of the order $1/(|I_{\bf
 d}|d)$. Such complicated structure of Bloch band $\varepsilon_{0+} ({\bf K})$
 for $\chi=\pi/4$ and $I_{\bf a}=I_{\bf a}= 4.3 $ is shown
 in the contour plot Fig.\ref{Fig1}(d).

Finally let us consider the possible experimental test of the
suggested gapless  spectrum of Majorana
fermions(\ref{Eq:spectrumLattice}).

 In addition to the  variety of experiments proposed \cite{Experiments}.
 the electronic thermal conductivity $\kappa$
measurements have been proven as an effective tool to study the
quaiparticle spectrum in the vortex phase \cite{Lowell}. The
electronic states in magnetic Bloch bands
(\ref{Eq:spectrumLattice})
 can carry the energy current in the direction
$\perp {\bf B}$  due to the hopping of quasiparticles between
neighboring vortices. The electronic spectrum on Abrikosov lattice
(\ref{Eq:spectrumLattice}) is gapless even in the regular lattices
provided the intervortex distance is $d<d_c$. Hence one should
expect the threshold behavior of $\kappa (B)$ in the increasing
magnetic field in the limit $T\rightarrow 0$. That is $\kappa$
should be zero in the gapped phase and in the gapless regime it
can be estimated by the textbook expression
 $\kappa_\perp \sim T V_g^2\tau \nu$ where
  ${\bf V_g} = \partial \varepsilon/ \partial {\bf K} $ is the
  group velocity of Bloch waves (\ref{Eq:spectrumLattice}), $\tau$ is
  transport time and $\nu= \omega^{-1} d^{-2}$ is density of
   in the vortex lattice. The typical value of
  group velocity determined by the intervortex hopping is $V_g \sim V_F e^{-d/\xi} \sqrt{d/k_F\xi^2}$
 so that assuming $d=\beta\xi\sqrt{H_{c2}/B}$ where $\beta\sim 1$ we obtain
 \begin{equation}\label{Eq:ThCond}
\kappa_\perp/\kappa_N =  \frac{\sqrt{B/H_{c2}}}{\beta k_F\xi}
  e^{-2\beta\sqrt{H_{c2}/B}}.
 \end{equation}
   Besides that $\kappa_\perp$ also
 contains oscillating part due to rapid oscillations of energy levels (\ref{Eq:spectrumLattice}) with the
 period $k^{-1}_F$ by the intervortex distance. However the oscillations are mostly cancelled out due to the
 complicated structure of Bloch bands shown in
 Fig.\ref{Fig1}(d). The obtained value of $\kappa$ is valid in the superclean limit
 $\omega\tau>1$ for fully gapped superconductors including $s$ and $p_x+ip_y$ wave
symmetries. The estimation (\ref{Eq:ThCond}) contains small
prefactor $1/k_F\xi$ which can explain the experimentally observed
small values of $\kappa_\perp$ at $B\ll H_{c2}$ \cite{Lowell}.
Interestingly the observed\cite{ExpThCondKes} similar to
(\ref{Eq:ThCond}) behavior of $\kappa_\parallel$ in the direction
$\parallel {\bf B}$ can be explained by the theory \cite{MS2}
applied to the spectrum(\ref{Eq:spectrumLattice}) since the number
of conducting modes along vortex line is determined by the
tunnelling factors $I_{\bf d}$.

%The amplitude of the
% magnetic oscillations of $\kappa (B)$
% is given by $ \kappa \approx T V_F e^{-2d/\xi} \xi^2/d$.
%  This estimation holds at arbitrary low temperatures including
%  the quantum limit $T\ll \omega$
%   provided the intervortex distance is smaller than the critical
%  one $d<\xi \ln(k_F\xi)/2$. The discussed effect of heat conductivity oscillations in
%  magnetic field is analogous to the Shubnikov- de Haas
%  oscillations of electrical conductivity in normal metals and
%  we believe that it deserves further detailed investigation, both
%  theoretical and experimental.

 The author thanks Alexander Mel'nikov  and Sergei Sharov for many stimulating
 discussions and Ville Lahtinen for correspondence.
 The work was supported b y Russian Foundation for Basic Research.

\end{document}